\documentclass[11pt, draftclsnofoot, onecolumn]{IEEEtran}
\usepackage{cite,graphicx,amsmath,amssymb}
\usepackage{subfigure}
\usepackage{citesort}
\usepackage{fancyhdr}
\usepackage{mdwmath}
\usepackage{mdwtab}
\usepackage{balance}
\usepackage{xcolor}
\usepackage{bm}
\usepackage{amsthm}
\usepackage{algorithm}
\usepackage{algorithmic}
\usepackage{multirow}
\usepackage{flafter}
\usepackage{threeparttable}

\newcommand{\tabincell}[2]{\begin{tabular}{@{}#1@{}}#2\end{tabular}}

\begin{document}
\title{Artificial Intelligence Driven UAV-NOMA-MEC in Next Generation Wireless Networks}
\markboth{\textit{A Manuscript Submitted to The IEEE Wireless Communications} }{}

\date{\today}
\author{
Zhong~Yang,~\IEEEmembership{Student Member,~IEEE,}
Mingzhe~Chen,~\IEEEmembership{Member,~IEEE,}
Xiao~Liu,~\IEEEmembership{Student Member,~IEEE,}
Yuanwei~Liu,~\IEEEmembership{Senior Member,~IEEE,}
Yue~Chen,~\IEEEmembership{Senior Member,~IEEE,}
Shuguang~Cui,~\IEEEmembership{Fellow,~IEEE,}
H.~Vincent~Poor,~\IEEEmembership{Fellow,~IEEE}

\thanks{Z.~Yang, X.~Liu, Y.~Liu and Y.~Chen are with Queen Mary University of London, London,
UK (email: \{zhong.yang, x.liu, yuanwei.liu, yue.chen\}@qmul.ac.uk).}
\thanks{M.~Chen is with the Department of Electrical Engineering, Princeton University, Princeton, NJ 08544 USA, and also with the Shenzhen Research Institute of Big Data, The Chinese University of Hong Kong, Shenzhen 518172, China (e-mail: mingzhec@princeton.edu).}
\thanks{S.~Cui is with the Shenzhen Research Institute of Big Data, and the Future Network of Intelligence Institute, The Chinese University of Hong Kong, Shenzhen 518172, China (e-mail: shuguangcui@cuhk.edu.cn).}
\thanks{H. V. Poor is with the Department of Electrical Engineering, Princeton University, Princeton, NJ 08544 USA (e-mail: poor@princeton.edu).}
}
 \maketitle

\begin{abstract}

Driven by the unprecedented high throughput and low latency requirements in next generation wireless networks, this paper introduces an artificial intelligence (AI) enabled framework in which unmanned aerial vehicles (UAVs) use non-orthogonal multiple access (NOMA) and mobile edge computing (MEC) techniques to service terrestrial mobile users (MUs). The proposed framework enables the terrestrial MUs to offload their computational tasks simultaneously, intelligently, and flexibly, thus enhancing their connectivity as well as reducing their transmission latency and their energy consumption. To this end, the fundamentals of this framework are first introduced. Then, a number of communication and AI techniques are proposed to improve the quality of experiences of terrestrial MUs. To this end, federated learning and reinforcement learning are introduced for intelligent task offloading and computing resources allocation. For each learning technique, motivations, challenges, and representative results are introduced. Finally, several key technical challenges and open research issues of the proposed framework are summarized.
\end{abstract}

\vspace{-0.3cm}
\section{Introduction}

In next generation wireless networks, the stringent delay requirements of services and applications, such as virtual reality, augmented reality, holographic telepresence, industry 4.0, and robotics, are considerably restricted by finite battery and computing resources of terrestrial mobile users (MUs) and terrestrial access ponits (APs). In order to satisfy these stringent requirements, novel highly efficient techniques, such as mobile edge computing (MEC)~\cite{Zhou2019commmag}, non-orthogonal multiple access (NOMA)~\cite{Yuanwei2017pieee}, unmanned aerial vehicles (UAVs)~\cite{zhang2019tcom,yang2019twc}, and artificial intelligence (AI) algorithms~\cite{Zhang2020ieeenetworks} should be thoroughly investigated for next generation wireless networks.

In this light, early research articles have studied these techniques to effectively exploit the performance enhancement for next generation wireless networks. In~\cite{Zhou2019commmag}, fog computing is introduced for mobile networks which is capable of achieving higher capacity than conventional communication networks. The authors in~\cite{zhang2019tcom} investigate both cellular-enabled UAV communication and UAV-aided cellular communication and optimize the trajectory of the UAV subject to practical communication connectivity constraints. Reference~\cite{yang2019twc} minimize the sum energy consumption of MUs and UAVs in a UAV-MEC network by jointly optimize the user association, power control, computing resources allocation and location planning. A disaster resilient three-layered architecture is proposed in~\cite{Kaleem2019ieeenet}, in which UAV layers are integrated with edge computing to enable emergency communication links. In UAV-NOMA-MEC systems, a critical challenge is task offloading decision-making and computing resources allocation. Moreover, a natural approach to task offloading and computing resources allocation is to combine them. For this reason, they are often formulated as a mixed integer programming (MIP) problem~\cite{Wang2019tcom,Kiani2018iot}. In~\cite{Wang2019tcom}, the authors proposed a joint optimization approach to allocate both the communication resources and computing resources for NOMA-MEC networks, while minimizing the total energy consumption of MUs. The authors in~\cite{Kiani2018iot} minimize the energy consumption by adjusting the computing resources and transmit power of the APs.

MEC is a promising technique for next generation wireless networks, which moves the computing resources of central networks towards the network edges to MUs. MEC is capable of significantly improving the computing performance of MUs with low energy consumption. NOMA, with high bandwidth efficiency and ultra high connectivity, is an emerging technique in next generation wireless networks. In UAV-NOMA-MEC, NOMA is capable of enabling offloading multiple computational tasks simultaneously from a large number of MUs under stringent spectrum constraints. In UAV-NOMA-MEC systems, UAVs are equipped with computing capabilities, thus can be swiftly deployed to emergency situations when terrestrial MEC servers are overloaded or unavailable to MUs. There are two aspects to the combination of UAVs and communication, namely, UAV aided communications and communication for UAV operations. For the first aspect, UAV aided communication has been recognized as an emerging technique due to its superior flexibility and autonomy~\cite{yuanwei2019wcom}. For the second aspect, the operational control of the UAVs often relies on wireless communication, which introduces difficult challenges for spectrum allocation and interference cancellation.

With the rapid progression of artificial intelligence (AI) and the high-performance computing workstations, the integration of AI and UAV-NOMA-MEC is a promising direction to obtain an efficient joint resource allocation solution in an intelligent fashion. Firstly, deep reinforcement learning (DRL) is a model-free solution for efficient decision-making problems, such as task offloading decision and computing resources allocation in UAV-NOMA-MEC systems. Then, the distinguished fitting qualification of deep neural networks (DNNs) is a novel approach to predict the computational tasks in UAV-NOMA-MEC systems, which can be used to further improve the performance of above-mentioned resources allocation solutions. Moreover, a recently proposed federated learning (FL) model is capable of further enhancing the training efficiency of the DRL and DNNs.

The above challenges motivate us to consider an AI enabled UAV-NOMA-MEC framework in this paper, the rest of which is organized as follows. In Section II, the system structure for the proposed UAV-NOMA-MEC framework is presented. In Section III, FL enabled task prediction for UAV-NOMA-MEC is investigated. The deployment design for UAV-NOMA-MEC is given in Section IV. AI enabled joint resource allocation for UAV-NOMA-MEC is presented in Section V, before we conclude this work in Section VI. Table I provides a summary of advantages and disadvantages of AI solutions for UAV-NOMA-MEC systems.

\begin{figure*} [htp]
\setlength{\abovecaptionskip}{-0.2cm}
  \centering
  \includegraphics[width=4.5in]{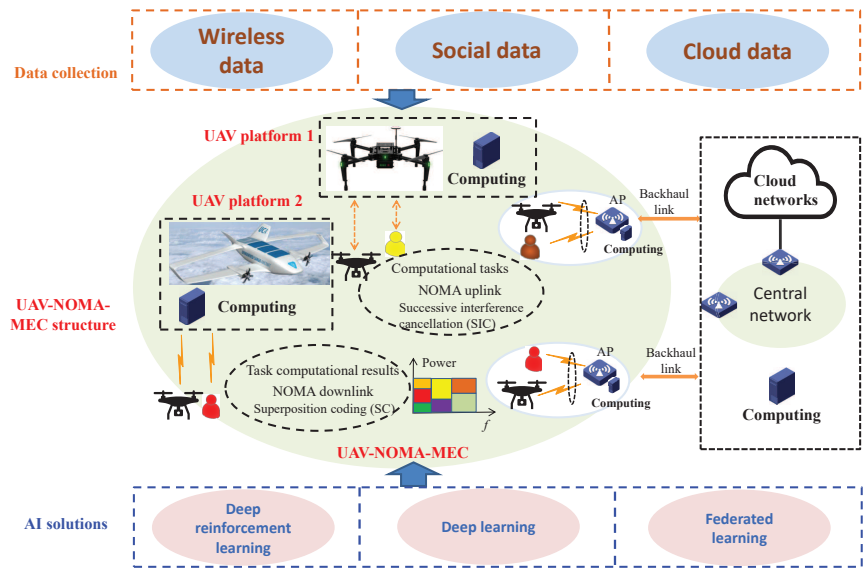}\\
  \caption{Network structure of UAV-NOMA-MEC}\label{structure}
\end{figure*}

\section{AI enabled UAV-NOMA-MEC System Structure}\label{section:nomaenablefrans}

\subsection{Structure for AI-enabled UAV-NOMA-MEC system}
Fig.~\ref{structure} illustrates the network structure of UAV-NOMA-MEC, which consists of a central network, multiple UAV platforms with computing capabilities, multiple APs with computing capabilities, and mobile MUs. MUs can be mobile smart devices or UAV platforms. Each MU has a computational task, which must be processed within a time constraint. In Fig.~\ref{structure}, UAVs and other MUs are clustered as one group, which has heterogeneous channel condition, thus is appropriate for NOMA uplink transmission. The MUs communicate with the UAV platforms or APs using the NOMA techniques, while the communication between APs and the central network is also conducted by the NOMA techniques. MUs can offload the computational tasks to UAV platforms and the APs located within its communication range. Moreover, the mobility of MUs and UAVs have heterogeneous characteristics, which is challenging for resource allocation. To allocate the communication resources and computing resources efficiently, we need to predict the mobility of MUs' tasks. Then, based on the predicted task mobility, UAVs are deployed accordingly. Lastly, task offloading decision and computing resources allocation are implemented with the proposed AI algorithms.
\subsection{Task Mobility Prediction for UAV-NOMA-MEC}

Due to the mobility characteristic of MUs and computational tasks in UAV-NOMA-MEC networks, the requested computational tasks varies over time. Therefore, the computing resources allocation and the task offloading decision must be conducted dynamically according to the task mobility. To efficiently allocate computing resources in UAV-NOMA-MEC, some prior information is required, e.g., task mobility in the future. The recent advances in AI have provide novel approaches to predict the task mobility. The advantage of AI algorithms is that they can train a learning model to obtain the complex relationship between the future task mobility and the task mobilities in the history, which is non-trivial for conventional approaches. Therefore, we propose AI algorithms for task mobility prediction, which works as a prior information for joint resources allocation (e.g., bandwidth, storage capacity and computing speed, etc).
\begin{table*}[t!]
\caption{Advantages and Disadvantages of Artificial Intelligent Solutions for UAV-NOMA-MEC}
    \centering
	\begin{tabular}{|l||l||l|}\hline
    	 {\bf AI~solutions}&{\bf Advantages for UAV-NOMA-MEC}&{\bf Disadvantages for UAV-NOMA-MEC}\\\hline
         \tabincell{l}{Deep neural networks \\(DNN)} & \tabincell{l}{(a) distinguished fitting capabilities \\of task prediction \\ (b) complex non-linear relationships \\of task prediction} &  \tabincell{l}{(a) require large amount of labelled input/output \\ wireless data, social data and cloud data \\(b) over fitting problem for task prediction}  \\\hline
         \tabincell{l}{Deep reinforcement \\learning (DRL)} & \tabincell{l}{(a) does not need labelled training data \\ for resource allocation         \\ (b) similar with humans resource allocation experience} &  \tabincell{l}{large action space and state space with \\ the increasing number of UAVs, BSs, and MUs}  \\\hline
         \tabincell{l}{Federated learning \\(FL)} & \tabincell{l}{(a) privacy preserving for sensitive MUs \\in UAV-NOMA-MEC \\ (b) high training efficiency for task prediction } & \tabincell{l}{local network (UAVs network, BSs networks, \\ and MUs networks) failure  affects \\the global network} \\\hline
	\end{tabular}
	\label{tableai}
\end{table*}
\subsection{Techniques of UAV-NOMA-MEC based frameworks}

\subsubsection{UAVs for NOMA-MEC networks}

UAVs have attracted research attention from both academia and industry for next generation wireless networks, because UAVs are easy to be deployed in various scenarios to support services, such as rapid emergency communication response and accurate observation services. In these services, UAVs are deployed as relays to support MUs with line-of-sight (LOS) wireless channels. When the computing capabilities of APs and MUs are not enough for massive tasks computing, deploy computing resources quickly to MUs is a major challenge for NOMA-MEC networks. UAVs can be deployed dynamically according to the requirements of MUs, thus are an efficient complementary for NOMA-MEC networks. In the proposed UAV-NOMA-MEC networks, UAVs work from two aspects, i.e., UAVs as base stations and UAVs as users. From the UAVs as base stations aspect, UAVs are integrated with computing resources and can be deployed dynamically for emergency use. However, deploying UAVs at the UAV-NOMA-MEC networks is challenging and a large amount of recent works have studied the deployment problem. Furthermore, in contrast to conventional terrestrial BSs deployment, the UAV placement is no longer a 2D placement problem, it is actually a 3D placement problem. From the UAVs as users aspect, UAVs have computing-intensive tasks, which require a large amount of computing resources. Therefore, the UAVs can transmit the computational tasks to the MEC servers at the terrestrial AP using NOMA technique. Then after computing, the tasks' computing results are transmitted back to the UAVs using NOMA technique.

\subsubsection{NOMA for UAV-MEC networks}

For UAV-MEC networks, choosing suitable transmission mechanism for the computational tasks offloading is a key challenge for reducing the computing delay. Different from orthogonal multiple access (OMA) in UAV-MEC, NOMA can ensure that multiple computational tasks are offloaded from MUs to UAV platforms or terrestrial MEC servers within the same given time/frequency resource block (RB), which is capable of significantly reducing the computation latency of MUs. For this reason, we adopt NOMA in UAV-MEC networks to better utilize the capacity of the communication channel for computational tasks offloading, and consequently reduce the task computational latency for multiuser UAV-MEC networks.

\subsubsection{AI for UAV-NOMA-MEC networks}

The recent advances in AI offer promising approaches to tackle the new challenges in UAV-NOMA-MEC. For instance, in order to efficiently allocate the limited communication and computing resources in UAV-NOMA-MEC, deep learning (DL) oriented algorithms can be used to predict task popularity more accurately, going beyond conventional approaches~\cite{Ramesh2017aaai}. Furthermore, deep reinforcement learning (DRL) algorithms can be utilized to solve stochastic optimization problems, which may not be computationally feasible with conventional optimization approaches. In fact, the time complexity of optimal solutions for the joint resource allocation problems arising in UAV-NOMA-MEC increase exponentially with the number of involved variables (e.g., number of MUs, number of UAVs, etc.). To this end, we propose AI based solution for UAV-NOMA-MEC framework. Table~\ref{tableai} presents the advantages and disadvantages of the AI based solution for UAV-NOMA-MEC framework.
\subsection{AI-enabled UAV-NOMA-MEC Network Optimization for Serving Terrestrial Users and UAVs}

The considered UAV-NOMA-MEC framework contains several optimization problems, including task prediction, UAV deployment, user association, signal processing, and joint resource allocation. The predicted tasks work as prior information of MUs' requirements, which is used for the following optimizations. With the predicted requirements, the UAVs are deployed accordingly, with reinforcement learning (RL) approaches, since RL algorithms are suitable for UAVs deployment optimization. Then, how to associate the UAVs and terrestrial MEC servers is another critical problem. In the end, joint optimization of task offloading and computing resources allocation can be solved by the proposed DRL algorithms.

\section{Federated Learning Enabled User Prediction for UAV-NOMA-MEC}
In this section, we first explain why we need to use FL for computing resources allocation in the proposed framework. Then, we discuss the challenges of using FL for the proposed framework. Finally, we use an example to show the implementation of FL for optimizing computational and task allocation in the proposed framework.

\begin{figure*}[htbp]
  \centering
  {\subfigure[Dynamic user association in UAV-NOMA-MEC networks.]{\includegraphics[width=7cm]{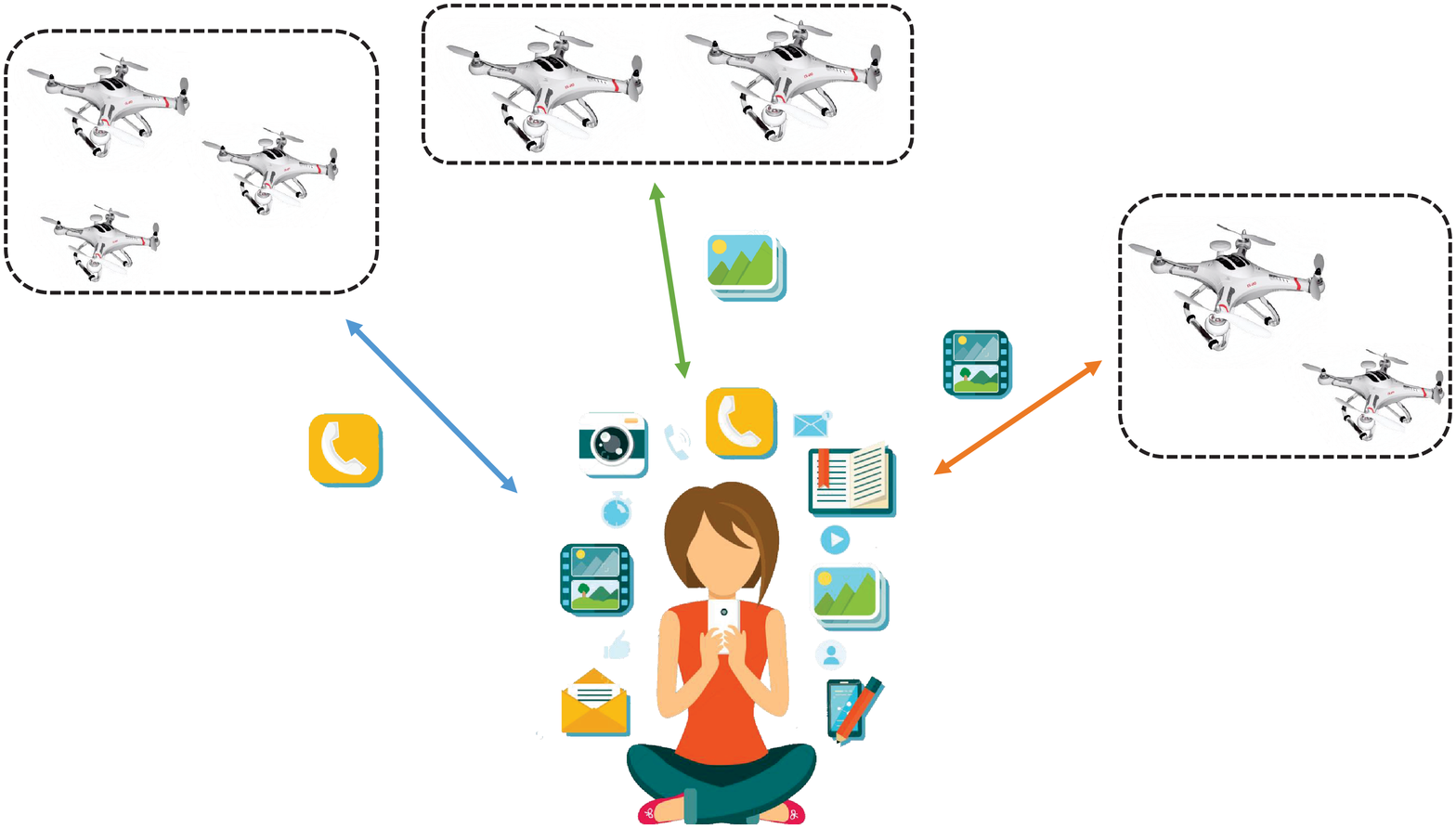}
\label{figUAVa}}
\subfigure[Implementation of FL over UAV-NOMA-MEC UAVs.]{\includegraphics[width=7cm]{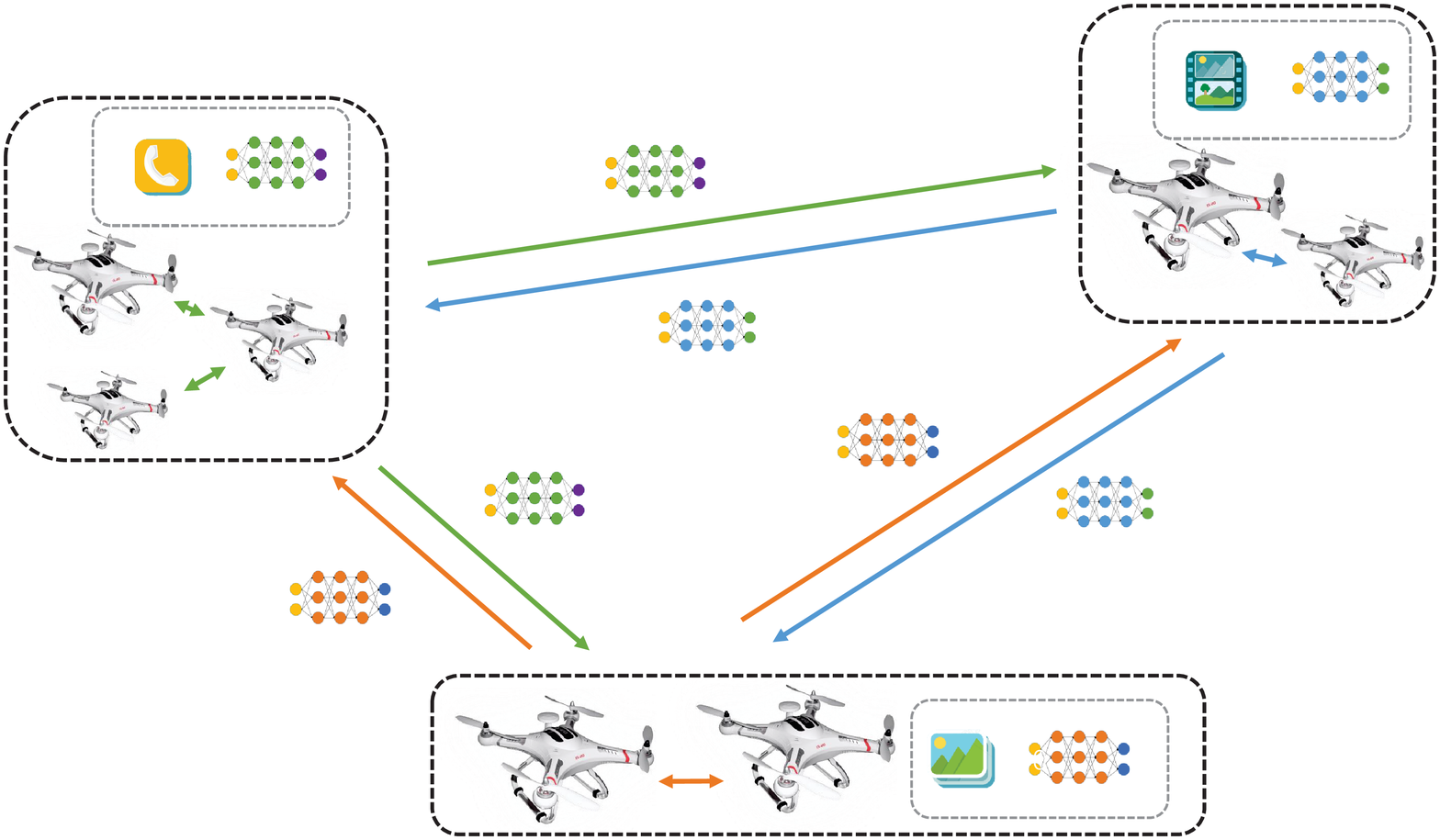}
\label{figUAVb}}
 }
  \caption{FL over UAV-NOMA-MEC Networks}\label{figUAV}
\vspace{-0.3cm}
\end{figure*}

\subsection{Motivations} Due to the mobility of UAVs and dynamic computational requests as shown in Fig. \ref{figUAVa}, a given terrestrial MU in the proposed framework must offload its computational tasks to different UAVs at different time periods. Hence, for a given terrestrial MU, its historical information related to computational tasks and user connection will be distributed across multiple UAVs. Due to the large data size of each user's historical information, the limited energy of each UAV, and privacy concerns, UAVs may not be able to share each MU's historical information with other UAVs. Therefore, it is necessary to design a novel distributed machine learning (ML) algorithm that enables the UAVs to train a common ML model that can accurately learn the entire future computational requests of each MU without data exchange. FL is a such type of distributed learning algorithm which enables the UAVs to exchange their trained ML parameters to generate a common ML model for computational request predictions \cite{9210812}. In particular, FL is trained by an iterative distributed process. At each iteration, each UAV must first use its collected data to train its local ML model and share its ML parameters to other UAVs, as shown in Fig. \ref{figUAVb}. Then, each UAV will use the ML parameters received from other UAVs and its own ML parameters to generate a common ML model. After that, each UAV will use the generated common ML model to update its own local ML model. After several iterations, each UAV can find a common ML model that can predict the entire future computational task requests of each MU.

\subsection{Challenges} Implementation of FL over UAVs also faces several challenges. First,  due to their mobility and limited battery capacity, UAVs may not be able to participate in all FL training iterations. In particular, the number of UAVs that can participate in different FL iterations is different thus affecting FL training loss. To this end, a UAV scheduling scheme must be designed so as to minimize the FL training loss and convergence time. Meanwhile, due to their high flying speed, UAVs must complete the FL training process in a strict time period. Hence, an efficient FL training method is needed for UAVs. In addition, UAVs in the proposed framework must provide computational services for the terrestrial MUs. Therefore, UAVs must find a tradeoff of the energy consumption for FL training and providing computational services for terrestrial MUs. Finally, in the proposed framework, each UAV must communicate with both terrestrial MUs and other UAVs. The mutual interference between such aerial and terrestrial systems will significantly affect the data rate of computational task transmission and ML parameter transmission thus increasing the FL convergence time and computational time. In consequence, there is a need to jointly optimize wireless resource allocation and deployment for UAVs so as to minimize the mutual interference.

\begin{figure*}[htbp]
  \centering
  {\subfigure[Prediction accuracy as the number of MUs varies.]{\includegraphics[width=7cm]{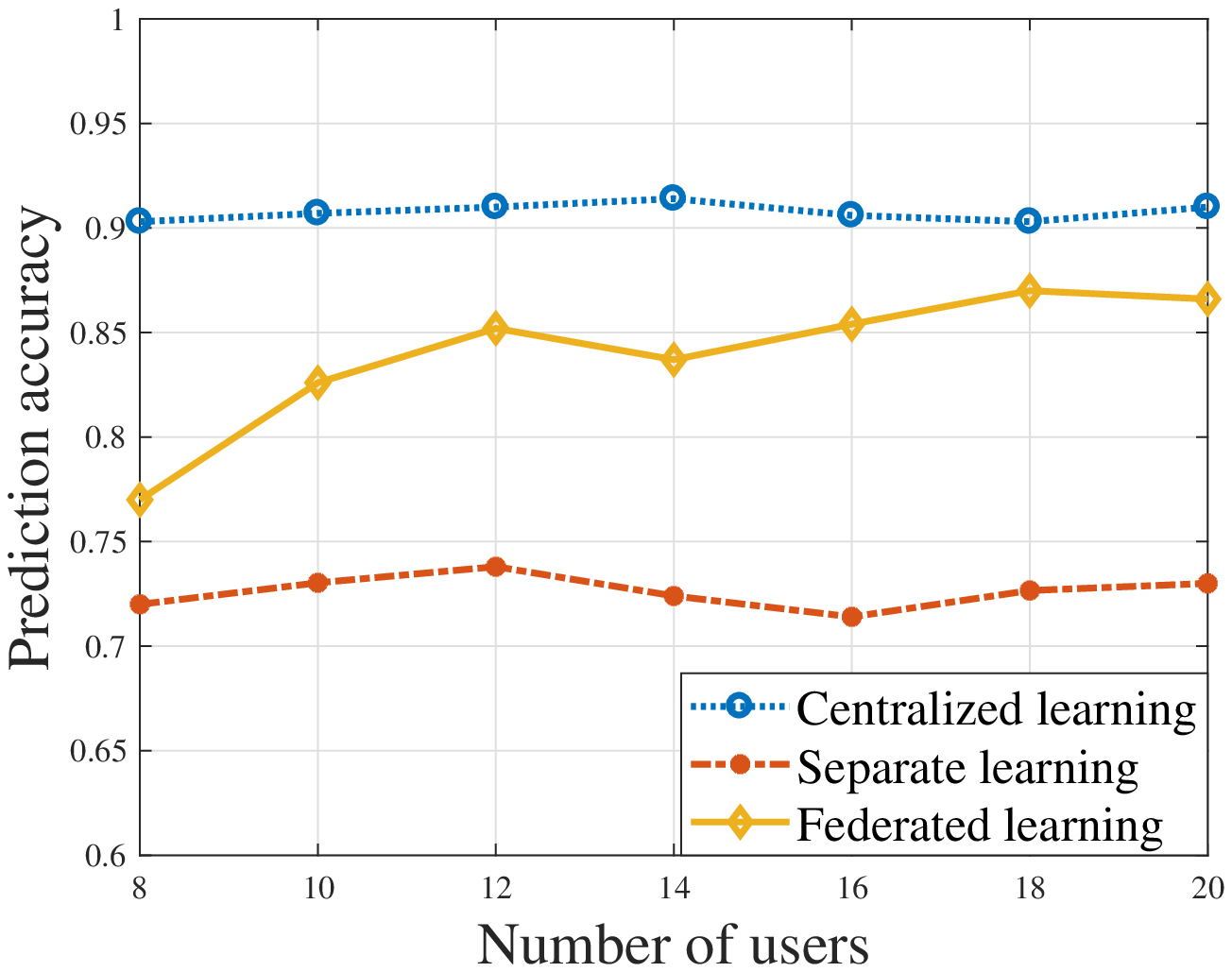}
\label{figFLa}}
\subfigure[Predicted user association as the data size of computational tasks varies.]{\includegraphics[width=7cm]{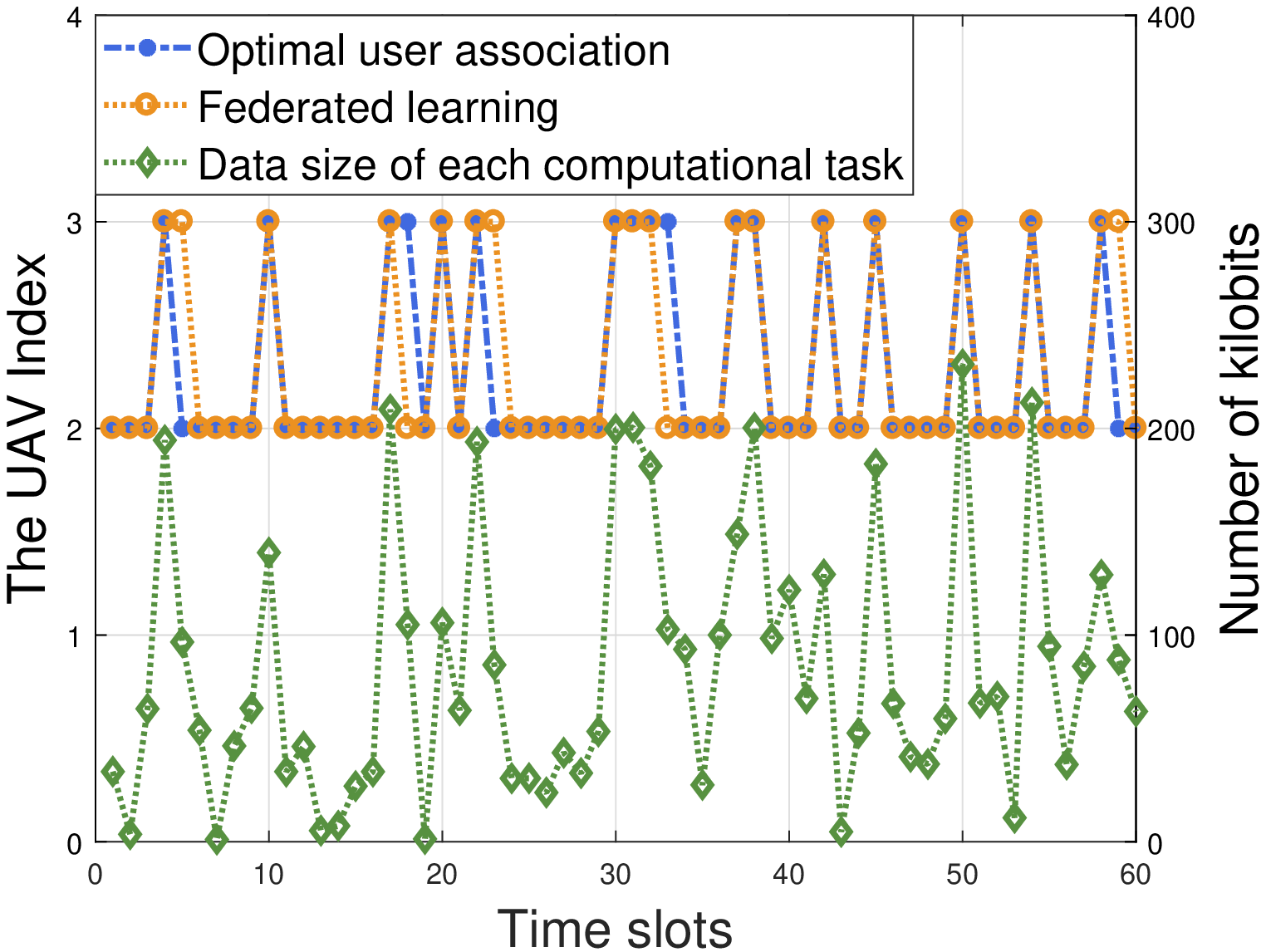}
\label{figFLb}}
 }
  \caption{FL for Proactive User Association}\label{fig:FLresult}
\vspace{-0.3cm}
\end{figure*}

\subsection{Representative Result} Next, we use two simulation figures to show the performance of using FL for the proposed framework. The simulation settings is based our previous work \cite{wang2020federated}. In particular, Fig.~\ref{fig:FLresult} shows the performance of using FL for proactively determining user association. Given the future user association, one can use optimization theory to optimize the task allocation and resource allocation. From Fig. \ref{figFLa}, we can see that FL can achieve a better accuracy compared to separate learning. This is because FL enables UAVs to cooperatively generate a common ML model and hence improving prediction accuracy. Meanwhile, as the number of MUs increases, the gap between centralized learning and FL decreases. However, different from centralized learning that requires UAVs to share their data, FL only needs the UAVs to share their learned ML parameters thus improve data privacy for the UAVs. Fig. \ref{figFLb} shows how the predicted user association changes as the data size of computational tasks varies. From this figure, we can see that FL can accurately determine the user association as the data size of computational tasks varies. This is because user association variable is binary and hence small FL prediction errors may not significantly affect the accuracy of the optimal user association prediction.

\section{Deployment design for UAV-NOMA-MEC}

\begin{figure*} [htp]
\setlength{\abovecaptionskip}{-0.2cm}
  \centering
  \includegraphics[width=4in]{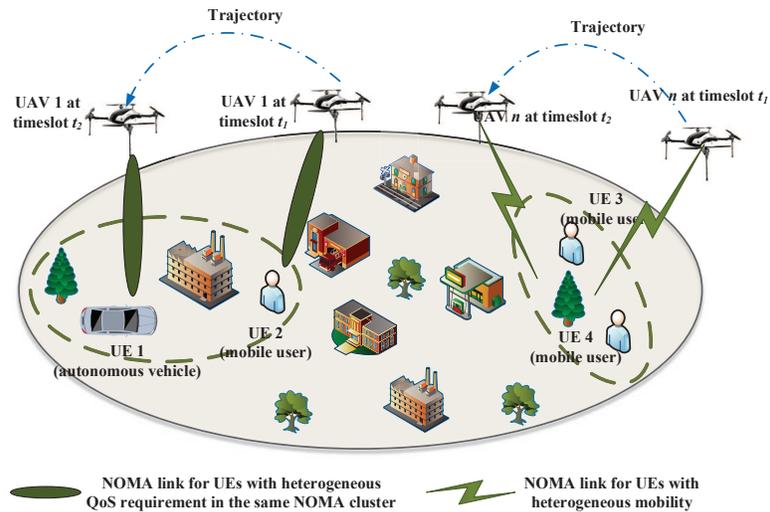}\\
  \caption{Air-ground integrated NOMA-MEC networks}\label{deployment}
\end{figure*}

As mentioned above, heterogeneous network segments, including heterogeneous user mobility, tele-traffic demand, and computing resource requirements, impose significant challenges on conventional terrestrial MEC networks. In an effort to tackle these challenges, the terrestrial MEC networks may be intrinsically amalgamated with UAV-aided wireless networks for forming air-ground integrated mobile edge computing networks. Compared to conventional terrestrial NOMA-MEC networks, UAVs can be dynamically deployed closer to MUs than terrestrial APs, which leads to improved performance. Additionally, as shown in Fig.~\ref{deployment}, in the NOMA-MEC networks, dynamic deployment design of UAVs is capable of making the channel condition of NOMA MUs more suitable to NOMA policy, which improves the system performance. In contrast to the UAV-aided wireless networks, where UAVs act as aerial base stations or aerial relays to improve the QoS of MUs, by integrating UAVs into NOMA-MEC networks, UAVs can act as aerial MEC server to execute computational tasks from MUs. Since heterogeneous user mobility is considered, UAVs has to be re-deployed simultaneously based on the movement of MUs. Moreover, when a particular UE has special request, UAVs can adaptively changing their positions to satisfy MUs' requirement. Moreover, as illustrated in Fig.~\ref{deployment}, in the NOMA-MEC networks where delay-sensitive MUs and delay-tolerant MUs are partitioned into the same cluster, UAVs can fly towards to delay-sensitive MUs to minimize the sum delay of all MUs. In this section, we discuss the deployment and trajectory design for UAV-NOMA-MEC networks.
Based on the network structure of UAV-NOMA-MEC networks, we focus on the AI-based solutions for designing the deployment of UAVs, this is because that UAVs operate in a complex time-variant hybrid environment, where the classic mathematical models have limited accuracy. In contrast to the conventional gradient-based optimization techniques, RL approaches are capable of enabling UAVs to rapidly adapt their trajectories to the dynamic/uncertain environment by learning from their environment and historical experiences.

In the RL-empowered UAV-NOMA-MEC networks, RL model empowers agents to make observations and take actions within the environment, and in return, receive rewards. It possesses learning capability based on correcting mistakes over trial and aims for maximizing expected long-term rewards. Hence, RL algorithms outperform the conventional algorithms in terms of dynamic scenarios or interactive with environment. However, every approach conveys both advantages and disadvantages in variable scenarios of UAV-NOMA-MEC networks. RL models assume the formulated as a Markovian problem, which indicates that when the current state depends not only on the attained previous state, RL algorithms may fail to solve the problem. Additionally, when faced with simple scenarios, RL algorithms have no superiority due to the reason that the optimality of RL algorithms cannot be theoretically proved or strictly guaranteed.

The discussions of designing architecture of RL model in UAV-NOMA-MEC networks are listed as follows:

\begin{itemize}
\item \textbf{Distributed or Centralized:} The advantage of centralized RL model in UAV-NOMA-MEC networks is that the central controller (the base station or control center) has complete local information. Thus it enables each agent (UAV) to cooperate with each other and searching for optimal control policy collectively. However, the centralized design requires the accurate instantaneous channel state information (CSI). Additionally, in the centralized ML model for UAV-NOMA-MEC networks, the central controller requires each agent to share their states and actions while searching for the optimal strategies. The formulated problem has to be solved by updating control policy based on all agents' actions and states, which leads to increased complexity of the model. On the other hand, the aforementioned challenge can be solved by distributed RL model. However, incomplete local information may lead to performance loss. Additionally, the distributed model causes unexpected state change of neighboring areas and leads to the complicated situation of multi-agents competition.
\item \textbf{Continuous or Discrete:} RL algorithms can be divided into three categories, namely, value-based algorithms, policy-based algorithms, and actor-critic algorithms. When consider discrete position, value-based RL algorithms are more suitable for designing the trajectory of UAVs. However, when discrete trajectory design problem is coupled with continuous task/resource allocation problem, how to design RL model with both continuous state space and discrete state space is challenging.
\end{itemize}

The problem of UAVs' trajectory design is coupled with other problems such as task offloading and computing resource allocation, which will be discussed in the next sections. UAVs' trajectory design problem can be jointly tackled with the other problems by adopting the RL solutions introduced in this section. In terms of challenges in UAV-NOMA-MEC networks, before fully reaping the benefits of integrating UAVs into NOMA-MEC networks, some of the weaknesses of UAVs such as their limited coverage area, meagre energy supply, as well as their limited backhaul have to be mitigated.

\section{AI enabled joint resource allocation in UAV-NOMA-MEC}
\begin{figure*} [htp]
\setlength{\abovecaptionskip}{-0.2cm}
  \centering
  \includegraphics[width=4in]{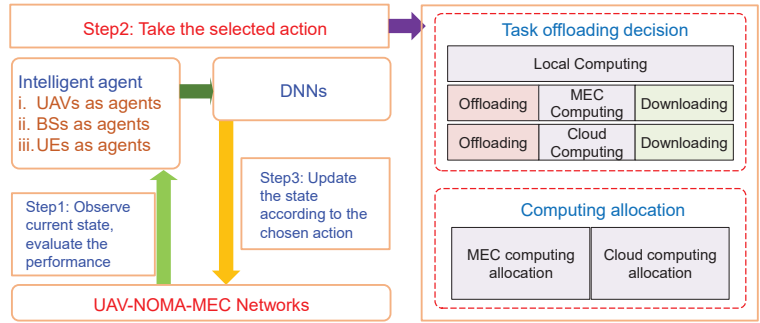}\\
  \caption{DRL based solution for joint resource allocation in UAV-NOMA-MEC}\label{structure3}
\end{figure*}

In this section, we advocate DRL based solutions for joint resource allocation of task offloading and computing resources allocation in UAV-NOMA-MEC networks. The structure of the DRL based solution is presented in Fig.~\ref{structure3}. The motivation of using DRL algorithms is to obtain an offline policy for the formulated joint optimization problem of task offloading and computing resources allocation.

\subsection{Joint task offloading and computing resources allocation in UAV-NOMA-MEC}

In multi-users UAV-NOMA-MEC networks, multiple MUs request for tasks computing services. The key research challenge is joint resources allocation, i.e., task offloading decision and computing resources allocation. More particularly, offloading computational tasks simultaneously to one destination, such as UAV and MEC server, is capable of reducing task computing latency. In UAV-NOMA-MEC, the task offloading decision and computing resources allocation are combined together, due to the reason that only the offloaded computational tasks need to be allocated with computing power from the computing platforms, such as UAVs and MEC servers. Therefore, we formulate the task offloading decision and computing resources allocation as a joint optimization problem.

In the proposed UAV-NOMA-MEC networks, tasks are offloaded simultaneously, using the NOMA technique, thus reducing the energy consumption of offloading and avoiding task transmission delay. Since noth the UAVs and MEC servers have computing capabilities, the task offloading in UAV-NOMA-MEC networks have more than one destinations. Further more, according to whether the computational tasks are segmented, there are two kinds of task offloading, namely, binary offloading and partial offloading.

\subsubsection{{ Binary offloading of UAV-NOMA-MEC}}

In the binary offloading of UAV-NOMA-MEC, the computational tasks are not segmented, so they are computed locally at MUs, or offloaded to UAVs and MEC servers for computing. So the task offloading decision for this case is to choose suitable destinations.

\subsubsection{{ Partial offloading of UAV-NOMA-MEC}}

In partial offloading of UAV-NOMA-MEC, the computational tasks are firstly divided into fragments. Then the offloading decision is to decide which fragment are offloaded to a specific destination, which is more complex than binary offloading.
\subsection{AI based solution for joint optimization in UAV-NOMA-MEC}

The prosperity of AI algorithms provide effective and low-cost solutions that make UAV-NOMA-MEC adaptive to the dynamic radio environment. We adopt RL in UAV-NOMA-MEC because the mechanical of RL algorithms is to obtain a long-term reward maximization by balancing exploration and exploitation, which is capable of solving a long-term optimization problem of joint task offloading and computing resources allocation~\cite{chaofan2019jsac,Sadeghi2019jsac}.

\subsubsection{Q-learning for joint optimization}
In UAV-NOMA-MEC, our objective is to obtain a offline policy for a long-term optimization of joint task offloading and computing resources allocation problem. Q-learning is one of the classic RL algorithms that is capable of selecting suitable action to maximize the reward in a particular situation by training the Q-table. The reward function of the Q-learning in UAV-NOMA-MEC is defined by the objective functions in the networks, e.g., energy consumption minimization, summation data rate maximization, computation latency minimization, etc. However, in Q-learning algorithm, the action selection scheme is based on a random mechanism, such as $\epsilon$-greedy.

\subsubsection{Modified reinforcement learning for joint optimization}

In RL algorithm, how to select the suitable action given the feedback and current state is critical. The action selection scheme is to balance the exploration and exploitation and avoiding over-fitting. Conventional $\epsilon$-greedy method cannnot balance the importance of current reward and future reward. Therefore, we proposed a Bayesian learning automata (BLA) based action scheme for the proposed modified RL algorithm in UAV-NOMA-MEC. The function of BLA is to adaptively make the decision to obtain the best action for the intelligent agent from the action space offered by the UAV-NOMA-MEC environment it operates in. It is proven that BLA based action selection scheme is capable of enabling every state to select the optimal action. The proposed BLA based RL algorithm achieves significant performance improvement against conventional RL algorithm in UAV-NOMA-MEC~\cite{Yang2020twc}.
\subsubsection{DQN in for joint optimization}
The dimensional curse of RL algorithms is a heavy burden for intelligent agent. Moreover, for UAV-NOMA-MEC, the dimensions of state space and action space are settled by the number of network parameters, e.g., number of channels, number of MUs and the number of MEC servers. To overcome this drawback, we adopt deep Q networks (DQN) for the joint optimization problem in UAV-NOMA-MEC. In the proposed DQN, the optimal policy of the intelligent agent is obtained by updating Q values in neural networks (NNs). The inputs of the NNs are the current states and the outputs are the probabilities of all the actions in the action space. By utilizing the fitting ability of the NNs, a high-dimension state input and low-dimension action output pattern is implemented to deal with the curse of dimensionality in conventional RL algorithms, especially when the number of network parameters in UAV-NOMA-MEC are large.
\section{Conclusion Remarks and Future Challenges}\label{section:Conclusion}
\subsection{Conclusion Remarks}
In this article, the design challenges associated with the application of AI techniques for UAV-NOMA-MEC networks have been investigated. An architecture for UAV-NOMA-MEC networks has been proposed, and key AI techniques for their optimization have been described. Then, the network structure of UAV-NOMA-MEC is demonstrated where the NOMA technique is adopted to accommodate multiple MUs in a single resource block. Furthermore, three specific techniques, namely, federated learning enabled task prediction, deployment design for UAVs, and joint resource allocation have been studied in detail.
\subsection{Future Challenges}
Although the advantages have been highlighted for task prediction, UAV deployment, and task computing in UAV-NOMA-MEC networks based on AI techniques, there still remain some open research issues and challenges to be addressed in the future, which are outlined as follows:
\begin{enumerate}
    \item[$\bullet$] {\bf{Combination with 6G Techniques:}} 6G provides significant new techniques that can be combined with UAV-NOMA-MEC, such as cell-free massive multiple-input multiple-output, millimeter-wave communication, and reconfigurable intelligent surfaces.

  \item[$\bullet$] { \bf{UAV trajectory and MA schemes selection:}} In UAV-NOMA-MEC, the UAV trajectory and multiple access (MA) schemes selection play a critical role in task offloading. AI based approaches can play an important role in jointly optimizing the UAV trajectory and MA scheme selection.

  \item[$\bullet$] { \bf{Joint optimization of AI transmission and wireless transmission:}} In AI algorithms, the network parameters need to be shared with other intelligent agents or network models. For AI enabled UAV-NOMA-MEC, the transmissions of network parameters in AI algorithms and wireless transmission need to be jointly optimized. A unified design of AI transmission and wireless transmission should be further investigated.

  \item[$\bullet$] { \bf{Joint optimization of UAVs, terrestrial MEC servers and MUs:}} A key aspect of the UAV-NOMA-MEC network is mobility of UAVs, terrestrial MEC servers and MUs, which brings significant challenge for the joint optimization of resource allocation. Therefore, more advanced approaches are needed to further explore the performance enhancement when all the elements are moving.

\end{enumerate}

\vspace{-0.3cm}
\bibliographystyle{IEEEtran}
\bibliography{mybib}

 \end{document}